\documentclass[prd,aps,floatfix,nofootinbib,twocolumn,tightenlines,showpacs]{revtex4}
\usepackage{dcolumn}
\usepackage{amsmath}
\usepackage{latexsym}
\usepackage{graphicx}
\usepackage{bm}

\def\k{\kappa}

\def\svev#1{\left\langle #1\right\rangle}       

\def\Tr{{\rm Tr}\,}

\long \def \blockcomment #1\endcomment{}

\def\Kth{\kappa_\text{th}}
\def\Kc{\kappa_c}

\newcommand{\bee}{\begin{equation}}
\newcommand{\ee}{\end{equation}}
\newcommand{\beea}{\begin{eqnarray}}
\newcommand{\eea}{\end{eqnarray}}

\begin{document}

\title{
Phase structure of SU(3) gauge theory with two flavors of symmetric-representation fermions
}
\author{Thomas DeGrand}
 \affiliation{Department of Physics,
University of Colorado, Boulder, CO 80309, USA}
\author{Yigal Shamir}
\author{Benjamin Svetitsky}
 \affiliation{Raymond and Beverly Sackler School of Physics and Astronomy,  Tel~Aviv University, 69978
Tel~Aviv, Israel}

\begin{abstract}
We have performed numerical simulations of SU(3) gauge theory coupled to
  $N_f=2$ flavors of symmetric representation fermions.
The fermions are discretized with the tadpole-improved clover action.
Our simulations are  done on lattices of length $L=6$, 8, and 12.
In all simulation volumes we observe a crossover from 
a strongly coupled confined phase to a weak coupling deconfined phase. 
Degeneracies in
screening masses, plus the behavior of the pseudoscalar decay constant, indicate that the deconfined phase is
also a phase in which chiral symmetry is restored. 
The movement of the confinement transition as the volume is changed is consistent with avoidance of the basin of attraction of an infrared fixed point of the massless theory.
\end{abstract}

\maketitle

\section{Introduction}
Many proposed extensions of the Standard Model involve new strong-coupling 
mechanisms that replace the fundamental, weakly coupled Higgs boson~\cite{Hill:2002ap}.
Among these are technicolor~\cite{Weinberg:1975gm,Susskind:1978ms} and  ``tumbling'' 
gauge dynamics~\cite{Raby:1979my}.
The dynamics of these pictures is based on weak-coupling ideas:
a perturbative $\beta$ function to evolve the coupling constant as the energy scale falls,
and most-attractive-channel arguments for scale separation
and selection of the condensed channel in tumbling.
Recently, groups have begun to test these proposals with nonperturbative lattice methods.
Most of the studies have been of SU(3) gauge theories with $N_f>3$ fundamental 
flavors~\cite{Brown:1992fz, Damgaard:1997ut, Iwasaki:2003de, Appelquist:2007hu,Deuzeman:2008sc,Fodor:2008hn,Jin:2008rc},
 but some \cite{Catterall:2007yx,Shamir:2008pb,DeGrand:2008dh,DelDebbio:2008zf,Catterall:2008qk,Hietanen:2008vc,Fodor:2008hm}
have considered other gauge groups and other representations for
the fermions, where a richer set of phenomena 
might exist~\cite{Kogut:1983sm,Kogut:1984sb,Kogut:1984nq,Kogut:1985xa,Karsch:1998qj}. 

Some phenomenological proposals depend on novel features of a given gauge theory's $\beta$ function.
One possibility is that the $\beta$ function has an infrared-attractive fixed point (IRFP) at 
finite gauge coupling~\cite{Caswell:1974gg,Banks:1981nn}. 
The infrared limit of the massless theory is then
scale-invariant and conformal, without confinement and without spontaneous breaking of chiral 
symmetry~\cite{Appelquist:1998xf,Appelquist:1998rb}.
Alternatively, the $\beta$ function might approach zero without actually vanishing, so that the 
running coupling is
nearly independent of scale over a wide range before confinement finally sets in at large distances.
This is the scenario of ``walking''~\cite{Bando:1987we,Cohen:1988sq}.
The $\beta$ function of the massless theory is thus an appealing handle for nonperturbative 
study~\cite{Appelquist:2007hu,Shamir:2008pb}.

In earlier work,
we began study of the SU(3) lattice theory with $N_f=2$ flavors of Wilson-clover fermions
in the sextet representation~\cite{Shamir:2008pb}.
Using the background field method (implemented
in lattice work as the Schr\"odinger 
functional~\cite{Sommer:1997xw}), we calculated a
discrete analogue of the $\beta$ function on a small lattice.
Contrary to perturbative estimates, we discovered that the discrete $\beta$ function vanishes
at a fairly weak value of the renormalized coupling.
If confirmed on larger lattices that allow extrapolation to the continuum limit, this would
constitute an IRFP, indicating a conformal IR theory.
The absence of spontaneous chiral symmetry breaking means that this theory
cannot be used for technicolor~\cite{Belyaev:2008yj}, but perhaps it can be incorporated in a theory of ``unparticles'' \cite{Georgi:2007ek}.

Since an IRFP implies unbroken chiral symmetry, 
the possibility of scale separation
cannot arise.
On the other hand, if the critical coupling for chiral
symmetry breaking is reached when the beta function is still negative,
the fermions will condense, acquire mass dynamically, and decouple
from lower energy scales.
Only the gluons will remain, and the theory will run towards confinement.
An early quenched study in the sextet theory \cite{Kogut:1984sb}
found large separation between the confinement scale and
a much higher chiral symmetry breaking scale, consistent with
the larger Casimir of the sextet representation.
When dynamical fermions are included,
lattice theories that are similar to ours---for example, QCD with
adjoint-representation fermions---do in fact show scale separation,
in that the critical temperature for
deconfinement is different from (and lower than) the critical temperature
for chiral symmetry restoration
\cite{Kogut:1984sb,Kogut:1985xa,Karsch:1998qj}.
This would imply, {\em inter alia\/}, that there is no IRFP in these theories.

Even if the massless theory possesses an IRFP, however, scale separation might reappear when one gives the
fermions a small mass.
Avoiding the IRFP, one can look for
a remnant of a chiral transition separated from the confinement transition.
Between the two transitions, one would find an intermediate phase where the
phenomenology of broken chiral symmetry would resemble that of QCD even
though the quarks are not confined.  More generally, it is interesting to see how the putative IRFP influences the physics of the massive theory and whether the latter can give evidence for or against the existence of the IRFP.

We therefore
present here a complementary study, based on straightforward calculations to
 determine the lattice theory's phase diagram and its particle spectrum \cite{DeGrand:2008dh}.
The salient points among our results are: (1) We find no scale separation, meaning that there is no intermediate phase and that deconfinement is coincident with chiral symmetry restoration; (2) the IRFP of the massless theory finds support in the behavior of the deconfinement transition as the size of the lattice changes.

The outline of the paper is as follows: In Sec.~\ref{sec:overview} we give a summary of our results
 by presenting a phase diagram of the lattice theory.
This diagram shows the critical curve $\Kc(\beta)$ where the quark mass,
defined through the Axial Ward Identity (AWI), vanishes; it shows as well the
 crossover/transition, associated with confinement and chiral symmetry, brought 
about by finite dimensions $L$ of the lattice.
We then proceed to a presentation of our work, beginning with the definitions of the
 theory and observables in Sec.~\ref{sec:theory} and continuing to detailed results in Sec.~\ref{sec:results}.
We offer further discussion in Sec.~\ref{sec:last}.

\section{Overview \label{sec:overview}}

We performed simulations on lattices of various sizes and measured both global
observables---plaquette, Polyakov loop---and spectral quantities---static
potential and hadronic correlators. 
If only the time dimension of the lattice%
\footnote{This is distinguished from the spatial directions by the antiperiodic 
boundary condition on the fermion field.} 
$N_t$ is finite and equal to $L$, then $(aL)^{-1}$ can be thought of as a nonzero temperature;
if each of the spatial dimensions of the lattice $N_s$ is equal to $L$, then $(aL)^3$ is a 
finite 3-volume.

\subsection{The deconfinement transition}
\begin{figure}
\begin{center}
\includegraphics[width=\columnwidth,clip]{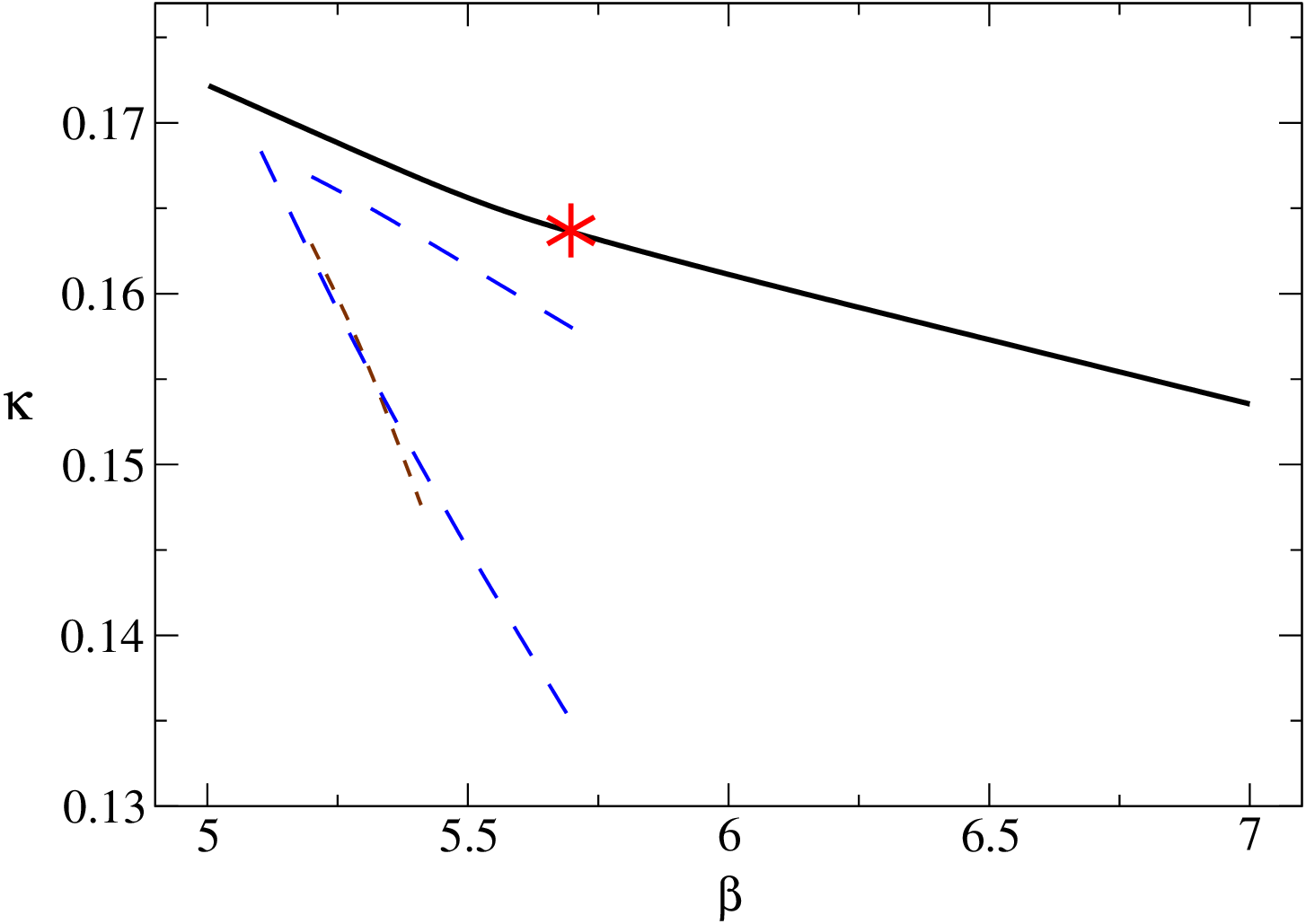}
\end{center}
\caption{Phase diagram  in the $(\beta,\kappa)$ plane.  The solid curve is
$\Kc(\beta)$, where $m_q$ vanishes; the dashed curves are
$\Kth(\beta)$, the thermal confinement transition, for three values of $N_t$: short dashes for $N_t=6$, long dashes for $N_t=8$ (lower curve) and $N_t=12$ (upper curve).
The star on the $\Kc$ curve marks the approximate location of the IR fixed point found in
 Ref.~\cite{Shamir:2008pb}.  The uncertainty in $\Kth$ is in the neighborhood of 0.001.
\label{fig:phase1}}
\end{figure}

We present in Fig.~\ref{fig:phase1} the phase diagram we have 
determined for the lattice theory in terms of the bare gauge coupling $\beta$
and the hopping parameter $\kappa$.
The solid curve is $\kappa_c(\beta)$, where the quark mass $m_q$ vanishes
 (see Sec.~\ref{sec:theory} for its definition).  $\kappa_c$ is, in principle, a 
feature of the theory in infinite volume, though of course we have determined it on a finite lattice.

The other (dashed) curves in Fig.~\ref{fig:phase1} 
refer to the location of finite-temperature transitions,
from confinement at small $\beta$ to non-confinement at large $\beta$.
There is one curve for each $N_t=6$, 8, and~12.
Data taken on volumes $N_s^3\times N_t=12^3\times 6$, $12^3\times 8$, and $12^3 \times 12$
 clearly show crossover behavior from a strong-coupling confined phase
to a deconfined phase, as
observed through the behavior of the Polyakov loop. 
The transition curves for different $N_t$ appear to approach the $\kappa_c(\beta)$ curve
to meet it near the same point.
The nearest we have come to this point is
 $(\beta,\kappa)=(5.1,\;0.169)$, which is still 
below $\k=\Kc$.
Simulation of the theory for stronger couplings is more difficult and we have not ventured into
 the $\beta<5.1$ regime.

Degeneracies among screening masses (see below) tell us that the deconfined phase is also a
phase in which chiral symmetry is restored.
We do not find two separate phase transitions for confinement and for chiral
symmetry breaking.

\begin{figure}
\begin{center}
\includegraphics[width=.9\columnwidth,clip]{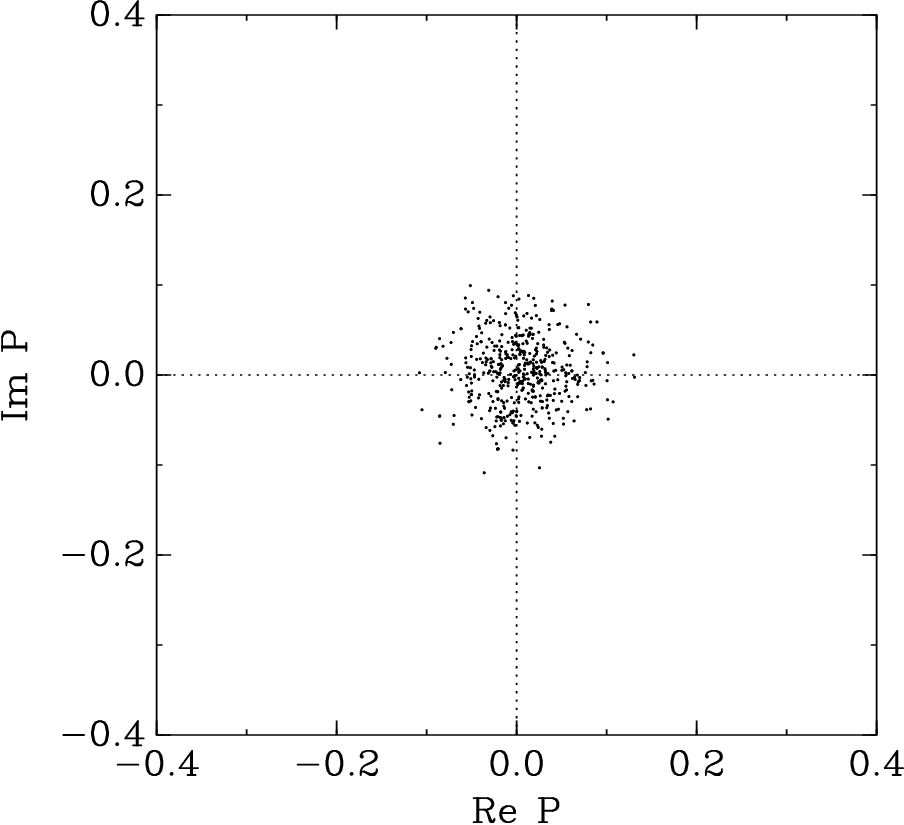}\\
\includegraphics[width=.9\columnwidth,clip]{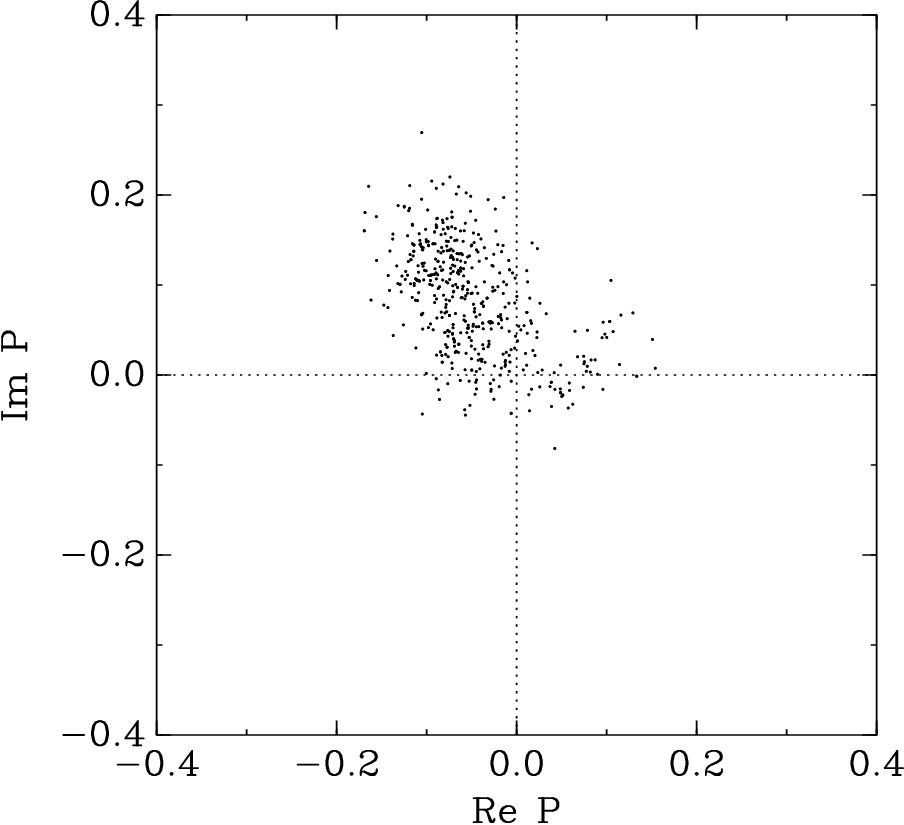}
\end{center}
\caption{Scatter plots of real and imaginary parts of the Polyakov loop from 
simulations at $\beta=5.5$, volume $8^4$ lattices.
Top: $\kappa=0.140$, in the confined phase.  Bottom: $\kappa=0.150$, in the deconfined phase.
\label{fig:scatter5.7}}
\end{figure}

The finite-temperature transition might be a true phase transition, rounded by
the less-than-infinite spatial volume, or it might be only a crossover.
In QCD with fundamental fermions, we expect this line to be a line of first order phase transitions
for heavy quark mass (small $\kappa$), terminating in a second order critical point and becoming only
a crossover as the quark mass falls. This is because fermions break the center $Z(3)$ symmetry,
and fundamental-representation fermions favor ordering the Polyakov loop along
the positive real axis.
Our data, however, indicate that sextet fermions break the center symmetry in a way that disfavors this direction
and favors $\text{arg}\, P=\pm2\pi/3$. An example of this behavior is shown in 
Fig.~\ref{fig:scatter5.7}.
This means that charge conjugation is spontaneously broken in the high temperature phase.
Hence, the crossover from the confined to the deconfined phase will become a
true phase transition
in the infinite volume limit.

We have marked on the $\kappa_c$ curve the approximate location of the IRFP found in the massless
 theory, as we reported in Ref.~\cite{Shamir:2008pb}.  There its location was determined by
 comparing the Schr\"odinger-functional effective coupling on lattices with $4^4$ and $8^4$ sites.
If it is a true IRFP then its location will approach a limit for
 sufficiently large values of $L$.  It is 
the IRFP that
 has caused us to focus our attention mainly on the region $5.1\leq\beta\leq6.0$.

The existence of an IRFP, not its precise location, is what is important.
The location
depends on the lattice action and on the particular scheme used to define
the renormalized coupling.
If the theory possesses
an IRFP, then there is a region in the space of bare coupling constants
that lies in its basin of attraction---a critical region. 
Correlation functions computed at values of
bare parameters in this basin will show scaling (power law) behavior at large
distance (plus, of course, cutoff-dependent behavior at short distance).

A quark mass breaks conformal invariance and so the critical region
 can include points only on the
$\Kc(\beta)$ line. Likewise, these points cannot be in the confinement phase:
confinement provides a dynamically generated infrared scale.
This means that lines of finite temperature transitions cannot intrude
into the critical region. This behavior is quite different from QCD with a small number of flavors,
where the deconfinement transition meets the $\Kc$ line at a point that moves
to ever larger $\beta$ as $N_t$ increases.
Thus the behavior of the deconfinement lines is an indirect signal for or against the
existence of scaling dynamics associated with an IRFP. 
The trend seen in Fig.~\ref{fig:phase1} is consistent with
the presence of such a critical region.

\subsection{Chiral symmetry restoration}
We have been able to distinguish two regions in the
 $\kappa<\Kc$ plane when $N_t$ is finite.  When the coupling is strong ($\beta$ small), there is confinement
 and spontaneous breaking of chiral symmetry.  At weak coupling chiral symmetry is restored (as
 much as possible for Wilson fermions) and there is no confinement.  

A phase with chiral symmetry breaking has a pseudoscalar mass $m_P$
that extrapolates to zero at zero
quark mass as $m_P^2\sim m_q$.
 This is not the same
as a strict proportionality $(am_P)^2\propto (am_q)$ along 
a line of fixed $\beta$,
because the lattice spacing itself will change as 
$\kappa$ changes.  Still,
at small quark mass the pion mass should become small compared to
all other dimensionful quantities, including the pseudoscalar decay constant
$f_P$.
All these other quantities will be controlled by the dynamically generated infrared scale and will remain nonzero in the chiral limit.

In a chirally restored phase we do not expect to see this mass hierarchy. Instead,
we expect to see parity doubling: the pseudoscalar and scalar 
 mesons should become degenerate,
as well as the vector and axial vector mesons. This effect is seen in
the screening masses measured in the high-temperature phase of ordinary QCD \cite{Born:1991zz}. In fact, we find that all four channels---scalar, pseudoscalar, vector,
axial vector---are close to degenerate.
This might be interpreted as a charmonium-like spectrum if the quarks are heavy.
If it persists as the quarks become massless then it indicates a weak quark--antiquark
interaction, perhaps one with no bound states.

A naive expectation for a screening mass is that it
behaves as
\bee
m_H^2 = 4\left [ \left(\frac{\pi}{N_t}\right)^2 + m_q^2\right]
\label{eq:minmat}
\ee
since ${\pi}/{N_t}$ is the lowest Matsubara frequency associated with
antiperiodic boundary conditions in a lattice of temporal length $N_t$. We observe all of this
behavior in the deconfined phase, and conclude that it is also a phase of chiral symmetry restoration.

\begin{figure*}
\begin{center}
\includegraphics[width=.9\textwidth,clip]{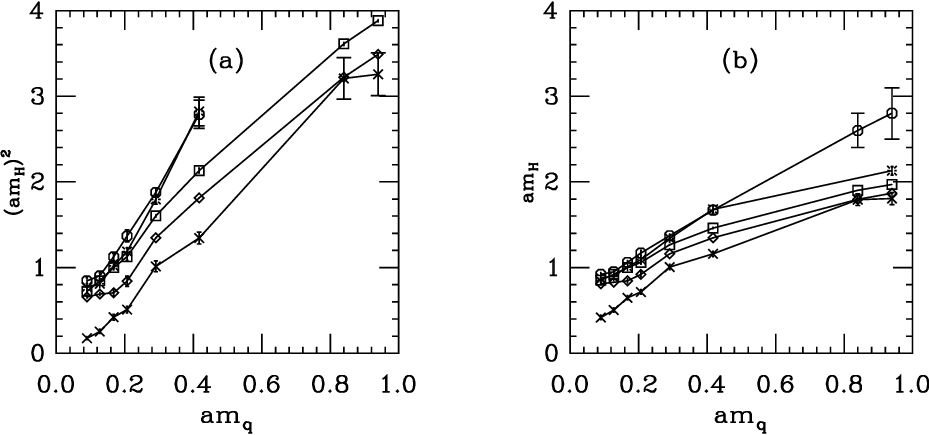}
\end{center}
\caption{ Screening masses and $f_P$
for $\beta=5.5$ on volume $(12\times 8^2)\times 8$.
In (a) we plot the squares of the quantities, while in (b) we plot the quantities themselves.
Crosses show $f_{P}$, pseudoscalars are diamonds, vectors are squares, octagons are axial vectors
 and bursts
are scalars.
\label{fig:mpi5.51288}}
\end{figure*}

We illustrate this behavior with results from a $(12\times 8)^2\times 8$ volume at $\beta=5.5$,
Fig. \ref{fig:mpi5.51288}. The points at $am_q>0.8$ are in the confined phase;
the other points are deconfined.
Even though $(am_P)^2$ appears to vary linearly with $am_q$
down to small $am_q$,
both $f_P$ and the degeneracy pattern point away from spontaneous chiral symmetry
breaking:
All states become degenerate and $af_P$ becomes small.

\section{Lattice action and simulation details \label{sec:theory}}
\subsection{Lattice action}

Our lattice theory is defined by the single-plaquette gauge action and a
Wilson fermion action with added clover term~\cite{Sheikholeslami:1985ij}.
We  modify the clover term's coefficient via tadpole improvement by setting $c_{SW}=1/u_0^3$.
We adopt the conventional self-consistent determination of the tadpole improvement factor, viz.,
\begin{equation}u_0^4= \frac13\svev{\Tr U_P}_{u_0},
\label{tadpole}
\end{equation}
where $U_P$ is the usual plaquette variable. 
This is the same action that we used in Ref.~\cite{Shamir:2008pb}.
In that paper we determined $\Kc$ on lattices with volume $4^4$, with Schr\"odinger-functional 
boundary conditions, and we fixed $u_0$ according to the space--space plaquette averages on 
those lattices at $\kappa=\Kc$ (see Table~\ref{Table1}).  Rather than recalculate 
$u_0$ at every $(\beta,\kappa)$, we choose to define the theory at every $\beta$ by fixing $u_0$ to 
take the value in Table~\ref{Table1}, irrespective of $\kappa$.

\begin{table}[t]
\caption{$\kappa_c$ and~$u_0$ as determined from  $L=4a$ simulations with Schr\"odinger functional boundary conditions
\cite{Shamir:2008pb}.
Linear interpolation may be used safely between $\beta=5.0$ and~5.5 and between $\beta=5.5$ and~6.0.
\label{Table1}}
\begin{ruledtabular}
\begin{tabular}{ccc}
$\beta$ & $\ \Kc\ $ & $\ u_0\ $ \\
\hline
5.0             &.1723  &.875\\
5.5             &.1654  &.887\\
6.0             &.1610  &.900\\
7.0             &.1536  &.916\\
8.0             &.1486  &.928\\
\end{tabular}
\end{ruledtabular}
\end{table}

\subsection{Data sets}

All simulations used the standard hybrid Monte Carlo algorithm. The trajectories in various 
runs were of lengths between 0.5 and 1.0, and the time steps ranged from~0.02 (at heavy quark 
masses) to~0.005 (for light masses).
The data sets at each of our $(\beta,\kappa)$ values consist of 300 to 1000 trajectories, 
with every fifth trajectory used for spectroscopy.

Results from five different simulation volumes are reported in this work:
\begin{itemize}
\item $12^3\times 6$ and
$12^3\times8$ are conventional finite-temperature lattices.
We use them to find  $\Kth(\beta)$ curves.
\item $(12\times 8^2)\times 8$ allowed faster runs than $12^3\times8$ and showed the same 
finite-temperature physics, though transitions are rounded by the smaller spatial volume. 
\item $8^3\times 12$ is a ``zero temperature'' lattice compared to $N_t=8$.
We use it to study how the {\em spatial\/} size $L=8$ intrudes on the $q\bar q$ 
potential and on meson masses.
\item $12^4$ has two roles.  One is as a ``zero temperature'' lattice, 
as long as we stay on the 
strong-coupling side of the $N_t=12$ confinement transition.  The other role is as 
a finite-temperature lattice that permits us to observe directly the movement of 
the $\Kc(\beta)$ curve when $N_t$ changes from 6 to 8 to 12. (We are well aware that
systematic thermodynamics studies require $N_s^3\times N_t$ lattices with $N_s \gg N_t$,
so we will be careful not to over-analyze our results.)
\end{itemize}
We also performed exploratory runs with $8^4$ volumes.

Dimensions of size 12 are where we determine meson masses.  If the dimension is temporal 
then the masses are conventional spectroscopic masses; if spatial then the masses are 
screening masses, affected by the Matsubara frequencies that create non-zero momentum 
transverse to the meson propagation.
Lattices where $N_t=12$ are also where we calculate the (spatial) $q\bar q$ potential from Wilson loops.

\begin{figure}
\begin{center}
\includegraphics[width=.9\columnwidth,clip]{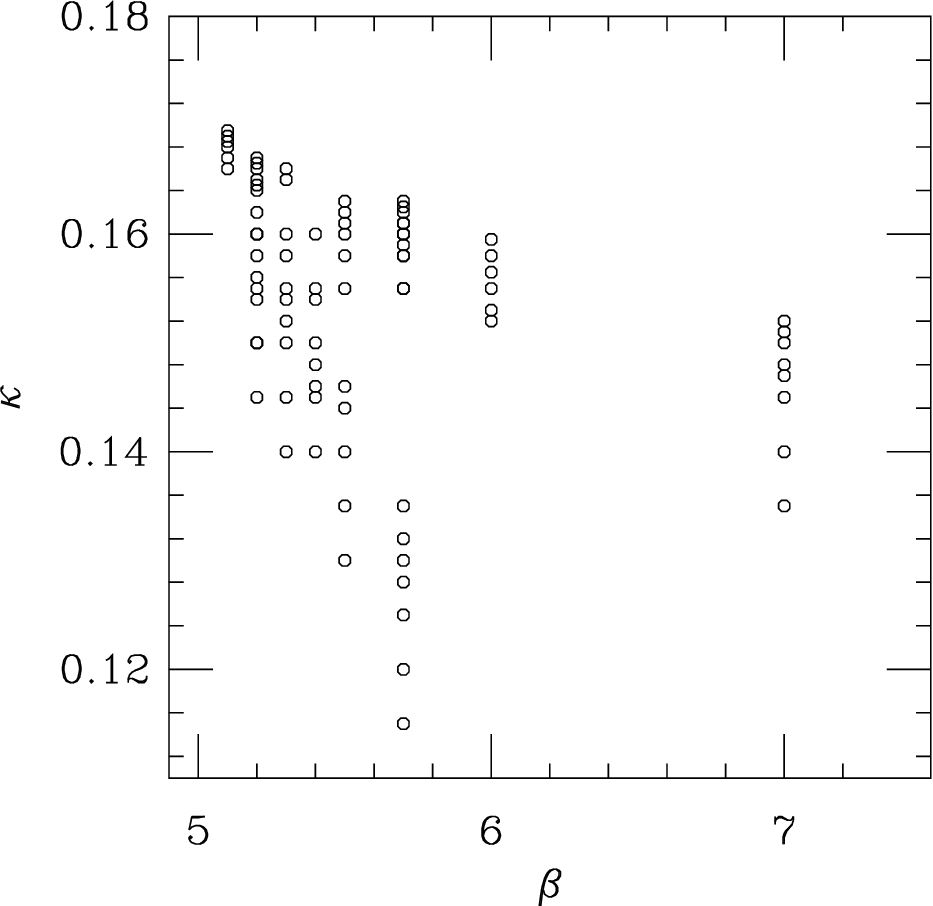}
\end{center}
\caption{Map  of our runs in the $(\beta,\kappa)$ plane. The upper set of points
parallels the $\Kc$ line in the deconfined phase. The lower set of points
populates the regions of  $\Kth$ lines.
\label{fig:runs}
}
\end{figure}

A map showing the bare parameters of our runs is shown in Fig. \ref{fig:runs}. We
concentrated on two areas of the $(\beta,\kappa)$ plane: The upper band of points
covers the area near $\kappa_c$, while the lower
band of points is in the vicinity of the deconfinement transitions.

\subsection{Meson propagators}
In order to measure masses when $N_t=12$, we effectively double $N_t$ with a trick that
has been used by several groups for computing weak matrix
elements \cite{Blum:2001xb,Aoki:2005ga,Allton:2007hx,Aubin:2007pt}. This combines periodic 
with antiperiodic boundary conditions as follows.
Take a valence Dirac operator with periodic temporal boundary conditions and
compute its propagator, $S_p(x)$ (we assume a source at $t=0$ for simplicity).
Take a second valence Dirac operator with antiperiodic temporal boundary
conditions, and compute its propagator $S_a(x)$. Now add the propagators to produce
\begin{equation}
S_{p+a}(x)=\frac{S_p(x)+S_a(x)}2\,,
\end{equation}
 and use this propagator
to construct meson correlators, e.g.,
\begin{equation}
C_{p+a}(t)=\int d^3x \svev{S_{p+a}(x)\gamma_5 S_{p+a}^\dag(x)\gamma_5}
\end{equation}
for a pseudoscalar meson. The resulting correlator will be a hyperbolic
cosine with midpoint at $t=N_t$.
(We use the same method in measuring screening masses when $N_s=12$.)

\subsection{Meson masses, decay constant, and quark mass}

After this, our fitting methodology is standard.
We perform spectroscopy using quark propagators computed in Coulomb gauge, generated from 
Gaussian wall sources.
We  take masses from correlated fits to long temporal ranges and use a combination 
of stable effective mass fits,
high confidence levels, and stability of our masses under variation of fit range to
 choose a result and its uncertainty.
As a general rule, when a meson  is heavy, single-exponential fits to correlation functions work 
well, but as the 
meson mass falls below $am_H \sim 1$ we need two-exponential fits to give a stable lightest mass.

We also measured the pseudoscalar decay constant and the AWI quark mass.
The latter is defined through
\bee
\partial_t \sum_x \svev{A_0(x,t)X(0)} = 2m_q \sum_x \svev{ P(x,t)X(0)}.
\label{eq:AWI}
\ee
where $A_0=\bar \psi \gamma_0\gamma_5 \psi$ and $P = \bar \psi \gamma_5 \psi$.
For consistency with the conventions of our Schr\"odinger functional calculation,
the derivative is taken to be the naive  difference 
operator $\partial_\mu f(x)=[f(x+\hat\mu) - f(x-\hat\mu)]/2$.
(We set $a=1$ in this discussion.)
The source $X(0)$ is arbitrary; for making the measurement as part of a spectroscopic
study
we use one made of our  Coulomb gauge Gaussian wave functions.
 When it is chosen to have pseudoscalar quantum numbers, and when we assume that
a single exponential dominates the correlator,
each
side of Eq.~(\ref{eq:AWI}) will be proportional to $\cosh(m_\pi (t-2N_t))$. We then extract the quark mass
by performing a three-parameter ($Z$, $m_{P}$, $m_q$) correlated fit to
\bee
\sum_x  \svev{ P(x,t)X(0)} = Z \left(e^{-m_{P} t} + e^{-m_{P}(2N_t-t)}\right)
\ee
and
\bee
\sum_x \svev{A_0(x,t)X(0)} =-\frac{ 2Z m_q}{\sinh m_{P}} \left(e^{-m_{P} t} - 
e^{-m_{P}(2N_t-t)} \right)  .
\ee
If we were to convert the lattice-regulated  AWI quark mass 
to some continuum regularization, it would require  multiplicative renormalization.
 We will neglect this overall renormalization 
factor in our discussion.
For Wilson fermions, radiative effects mix the local axial current $A_\mu=\bar \psi \gamma_\mu\gamma_5 \psi$
with $\partial_\mu \bar \psi \gamma_5 \psi$. 
The mixing is small in perturbation theory \cite{Luscher:1996vw}, so we neglect it as well.

We determine the pseudoscalar decay constant from the matrix element of the axial
vector current, $m_{P} f_{P} = \svev{ 0|A_0 | P }$. 
Specifically, we perform a three-parameter correlated fit to  the correlator of a Gaussian source and sink,
\bee
\sum_x  \svev{ X(x,t)X(0)} = \frac{Z^2}{2m_{P}} \left(e^{-m_{P} t} + e^{-m_{P}(2N_t-t)}\right),
\ee
and to the correlator of a Gaussian source and an axial current sink,
\bee
\sum_x \svev{A_0(x,t)X(0)} =\frac{ Z f_{P} }{2} \left(e^{-m_{P} t} - e^{-m_{P}(2N_t-t)}\right).
\ee

A lattice determination of a continuum decay constant
is a little involved. Because we are using
a non-chiral lattice action and a matrix element that does not precisely realize a Ward identity,
there is a lattice-to-continuum conversion factor $Z_A$  between the lattice matrix element
and a continuum-regulated decay constant,
\bee
f_{P}^{cont} = Z_A f_{P}.
\ee
We are aware of two ways to compute $Z_A$. One is nonperturbative, through the RI (regularization 
independent)
scheme. The other is
 through perturbation theory. For our exploratory study, we believe perturbation 
theory is adequate.
In the context of tadpole-improved perturbation theory \cite{Lepage:1992xa},
\bee Z_A = \left(1 + \frac{g^2}{16\pi^2}C_2(R) W\right)\left(1 - \frac{6\kappa}{8\kappa_c}\right) .
\ee
$C_2(R)$ is the appropriate Casimir (4/3 for fundamentals, 10/3 for sextets),
$W$ is a numerical factor
(5.79 for our tadpole-improved clover action),
and the ``tadpole factor'' $\left(1 - \frac{6\kappa}{8\kappa_c}\right)$
corrects the
field renormalization of lattice Wilson fermions compared to the continuum case.
Tadpole-improved perturbation theory  is designed to pick a scale for the coupling $g^2$ 
(the so-called
$q^*$ scale in its jargon). Again for our action, $q^*=2.41/a$. Computing the coupling from the 
plaquette
($g^2(q=3.41/a)=-3\ln \langle\Tr U_P/3\rangle$) and running to the desired $q^*$, we can estimate $Z_A$.
The bottom line is that the non-tadpole-improvement part of $Z_A$ (everything except the
$(1 - \frac{6\kappa}{8\kappa_c})$ factor) is essentially constant, equal to
about 0.8 over our parameter set.
For a quantity which is supposed to be  unity plus perturbative corrections,
the deviation is uncomfortably large.
If we were going to 
make phenomenological use of $f_\pi$---for example, to predict the technirho mass from a lattice measurement
of $m_\rho/f_\pi$ in a QCD-like theory---this would be cause for concern.
Since our only interest here is in whether $f_{P}$ vanishes in the chiral limit,
this is probably unimportant.

\subsection{The $\Kc$ curve \label{sec:Kc}}

The $\Kc$ curve shown in Fig.~\ref{fig:phase1} (see also Table~\ref{Table1}) was 
determined in \cite{Shamir:2008pb} by demanding $m_q=0$ on a lattice with $4^4$ sites 
and Schr\"odinger functional boundary conditions.
The matrix elements in the AWI (\ref{eq:AWI}) were evaluated at $t=2$ with the operator 
$X(0)$ defined through the boundary conditions themselves.
Finite-volume effects can shift $m_q(\kappa)$ and thus $\Kc(\beta)$.
We check this by calculating $m_q(\kappa)$ via the spectroscopic fit on $(12\times 8^2)\times 8$ and $8^3\times12$ lattices (Fig.~\ref{fig:mqvskappa}).
Apart from the data at strongest coupling ($\beta=5.1$), the
curves are quite smooth, and
on extrapolating to $m_q=0$ we
find  a shift upward of less than $2\times10^{-4}$ in $\kappa_c$, roughly the 
thickness of the solid curve in Fig.~\ref{fig:phase1}.

\subsection{Heavy quark potential}
In measuring the static potential between fundamental representation sources, our approach is, again,
fairly standard. We
extract $V(r)$ from the effective masses
of $R\times T$ Wilson loops after one level of HYP smearing~\cite{Hasenfratz:2001hp,Hasenfratz:2001tw}.
The short-distance effects of
the HYP smearing are corrected using a fit to the perturbative lattice
artifacts.
With fit parameters $A$, $B$, $\sigma$, and $\alpha$, we 
write $V(r)=V_\text{cont}+\alpha\Delta V_\text{latt}$, where
\bee
V_{cont}= \frac{A}{r} +  B + \sigma r
\ee
and
\bee
\Delta V_\text{latt} = V_\text{pert}(r)-\frac{1}{r}\ .
\ee
$\Delta V_\text{latt}$
is the difference between the exact lowest-order perturbative lattice propagator and $1/r$.
We quote values for the string tension $a^2\sigma$ and the Sommer parameter $r_0/a$ \cite{Sommer:1993ce}.
 For successful determination
of the potential
we demand consistent fit parameters from several different temporal sizes of the 
Wilson loops---fits to $T=t$ and $t+1$ for $t=3$, $4$, and perhaps~5, for example.

\section{Simulation results \label{sec:results}}

\begin{figure}
\begin{center}
\includegraphics[width=.9\columnwidth,clip]{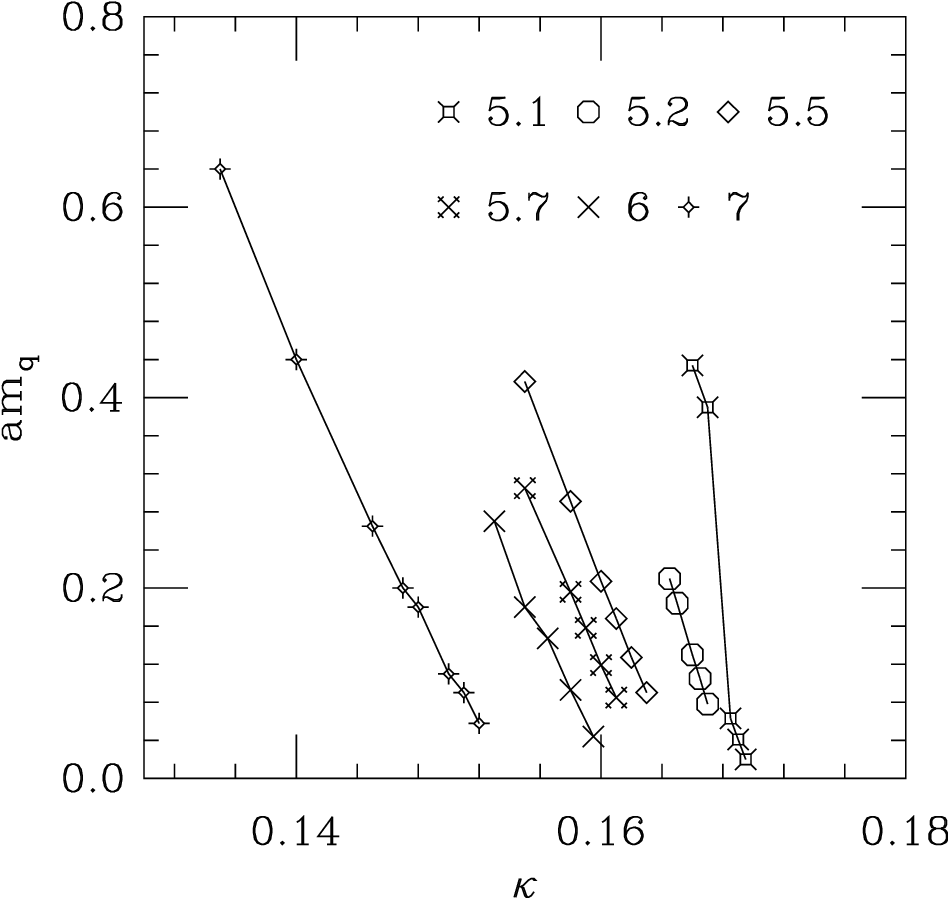}
\end{center}
\caption{AWI quark mass
from $(12\times 8^2)\times 8$ lattices. Curves are for (right to left) $\beta=5.1$, 5.2,
5.5, 5.7, 6.0, and~7.0.
\label{fig:mqvskappa}
}
\end{figure}

\subsection{Identifying the deconfinement line}
The straightforward way to find where the confinement scale crosses some distance $L$ is to find the 
finite-temperature confinement transition/crossover curve for $N_t=L$.
This we do via a conventional Polyakov loop analysis, looking for
metastability, phase coexistence, or a rapid variation in $\langle P\rangle$.
 Fig.~\ref{fig:phase1} shows the locations of the confinement transitions
for the three temporal sizes, $N_t=6$, 8, and 12.
We have data ranging from heavy to light quark masses. As we approach the $\Kc$ curve or make $\beta$ smaller than about 5.1, 
the number of conjugate gradient steps needed to compute the
fermionic force grows steeply. Hence we are unable to say whether the transitions actually join
the $\Kc$ line.

Measurements of the static potential on lattices with large $N_t$
agree  well
with Polyakov loop data.
In the confining phase, we use a confinement scale $R_c$ such as the Sommer
radius $r_0$ or $1/\sqrt{\sigma}$
(for string tension $\sigma$) to define a length scale.
Then lattices with a spatial size $L$ that is large compared to $R_c$ exhibit
confinement,
and those with $L\alt R_c$ do not. The ratio $R_c/a$
tends to grow as gauge coupling $\beta$ increases and as $\kappa$ rises.

\begin{figure*}
\begin{center}
\includegraphics[width=.9\textwidth,clip]{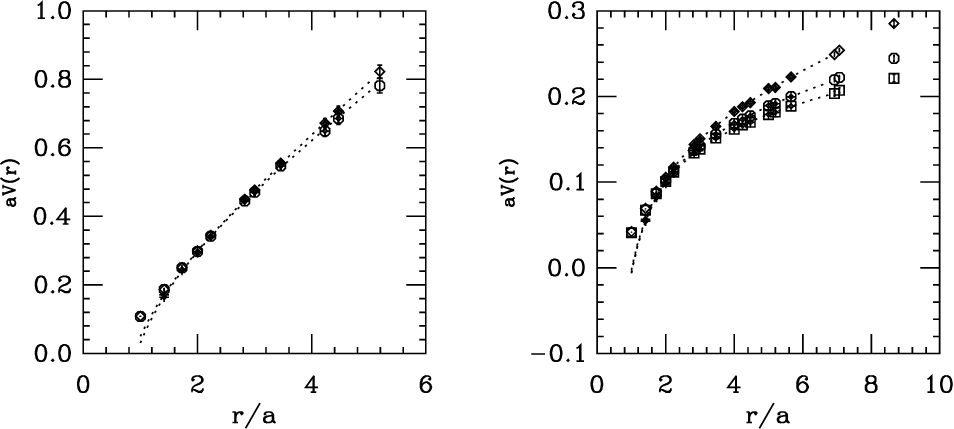}
\end{center}
\caption{Examples of potentials from our data sets. On the left, the string tension is
large, $r_0/a$ is
 not too small, and we can obtain a good fit to $V(r)$. On the right, the string tension has
 become small
and we cannot perform a reliable fit to $V(r)$.
Left panel, $8^3\times 12$,
$\beta=5.3$, $\kappa=0.155$.
Right panel, $12^4$, $\beta=5.7$, $\kappa=0.158$.
\label{fig:vt}
}
\end{figure*}

Examples of two fits to the potential
are shown in Fig. \ref{fig:vt}. On the left, we show results
from an $8^3\times 12$ lattice
at $\beta=5.3$, $\kappa=0.155$. The different plotting symbols denote two different temporal
sizes of the Wilson loops. This potential is clearly confining, with a large 
value of $a^2\sigma$ and small Sommer parameter $r_0/a$, either of which
can be used to define $R_c/a$.
Moving to larger $\beta$ and/or $\kappa$ will enlarge $R_c$ until confinement
will no longer be evident for $L=8$.

The right panel is from a $12^4$ lattice at $\beta=5.7$, $\kappa=0.158$. The potential has clearly
flattened: $a^2\sigma$ has become small and $r_0/a$ has grown. The different plotting symbols
show that our results are no longer independent of the temporal size of the
Wilson loop, and we are
unable to quote any number for $a^2\sigma$ or $r_0/a$.
Here, then, even $L=12$ cannot contain the confining region of the potential.

At very small $\beta$ we move deep into the strong coupling phase. The Sommer parameter $r_0/a$
becomes small and the fitter can no longer determine it.
The potential is purely linear, $V(r) \sim \sigma r$.
This means that, for all practical purposes,
there is a rather narrow region of bare parameters for which we can determine $a^2\sigma$ and $r_0/a$
from our simulation volumes. This region coincides roughly with the vicinity of the
$N_t=8$ deconfinement transition.

\subsection{Hadronic spectroscopy\label{sec:T0}}

As we have described, we use the behavior of $f_{P}$ and the $P$, $V$, $A$, and $S$ masses
to characterize the chiral symmetry aspects of each phase. 
If a decrease in the quark mass induces a drop in $m_{P}$ relative to the other masses 
while $f_{P}$
shows little variation, we call  the phase chirally broken.%
\footnote{For confirmation, we could then ask if $(m_{P}/\Lambda)^2 \propto m_q/\Lambda$ for some
choice of scale-setting mass $\Lambda$; but to do this successfully amounts to finding
a curve of constant physics.  See below.}
If on the other hand we observe parity doubling---degeneracy of the
$P$ and $S$ masses, and of the $V$ and $A$ masses---along with an $f_{P}$ that falls toward
zero along with $am_q$, we conclude that chiral symmetry is restored.

We expect the strong coupling phase to be a phase in which chiral symmetry 
is broken~\cite{Casher:1979vw}. 
We find that in this phase $am_{P}$, $am_{V}$, and $af_{P}$ are
easy to extract.
As most of our data are taken at fairly large quark masses, it is hard
to say more than that $am_{V}>am_{P}$.
The $P$ and $V$ masses fall as $\kappa$ grows.
The $S$ and $A$ signals are poor.
Their masses are large---greater than $am_V$---and the fits are
unstable.

The $P$ mass is not affected much by the deconfinement transition (apart from $\beta=5.1$).
Along most of the transition line, particle masses are large, $m_H a \sim 1$.

The $P$ and $V$ masses continue to fall as $\kappa$ is taken closer to $\Kc$ (i.e., as $am_q$ is taken to zero).
As we move into the deconfined phase, the signals in the $S$ and $A$ channels improve and their masses fall as well. All four masses---$P$, $S$, $V$, and~$A$---become nearly degenerate. At the same time,
$af_P$ becomes much smaller than any of the meson masses. As we mentioned
above, the simplest interpretation of our data is that chiral symmetry is restored in the deconfined phase.
The fourfold degeneracy may be reminiscent of the quark model with a weak hyperfine interaction, but another way to look at it is as a world where quarks form no bound states at all.  This, too, is characteristic of a massless limit with exact, unbroken chiral symmetry~\cite{Casher:1979vw}.

\begin{figure*}
\begin{center}
\includegraphics[width=.9\textwidth,clip]{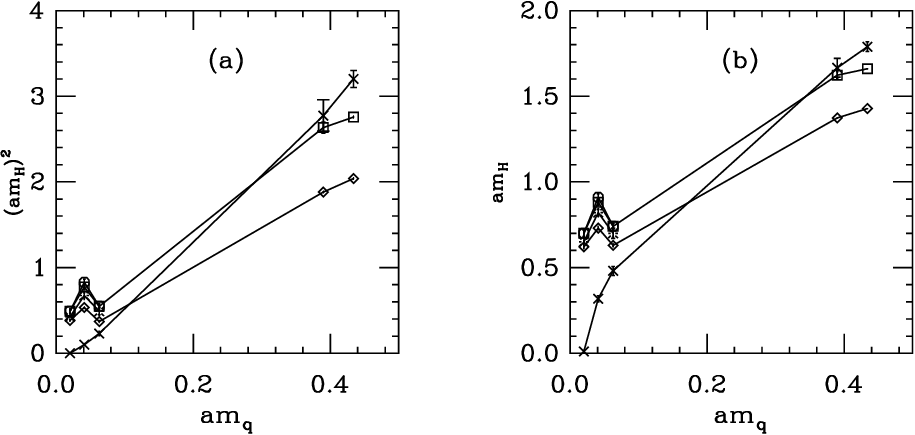}
\end{center}
\caption{ Screening masses and $f_P$
for $\beta=5.1$ on volume $(12\times 8^2)\times 8$.
Labels as in Fig.~{\protect{\ref{fig:mpi5.51288}}}.
\label{fig:mpi5.11288}}
\end{figure*}

\begin{figure*}
\begin{center}
\includegraphics[width=.9\textwidth,clip]{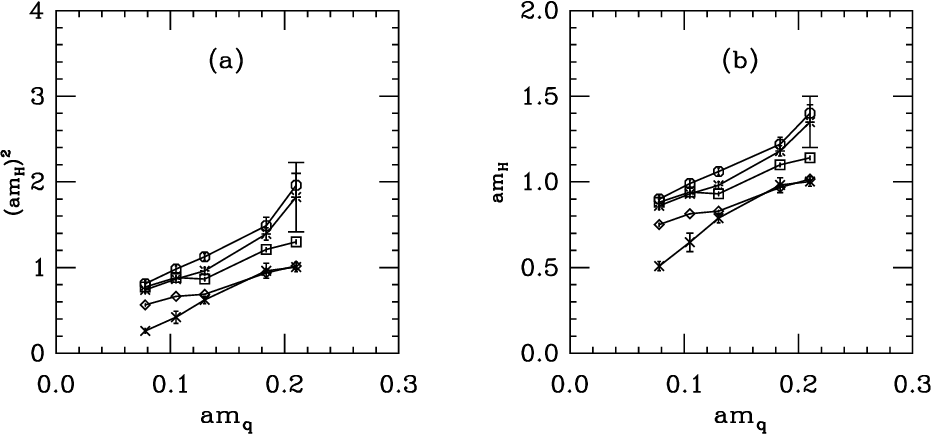}
\end{center}
\caption{ Screening masses and $f_P$ for $\beta=5.2$ on volume $(12\times 8^2)\times 8$.
Labels as in Fig.~{\protect{\ref{fig:mpi5.51288}}}.
\label{fig:mpi5.21288}}
\end{figure*}

\begin{figure*}
\begin{center}
\includegraphics[width=.9\textwidth,clip]{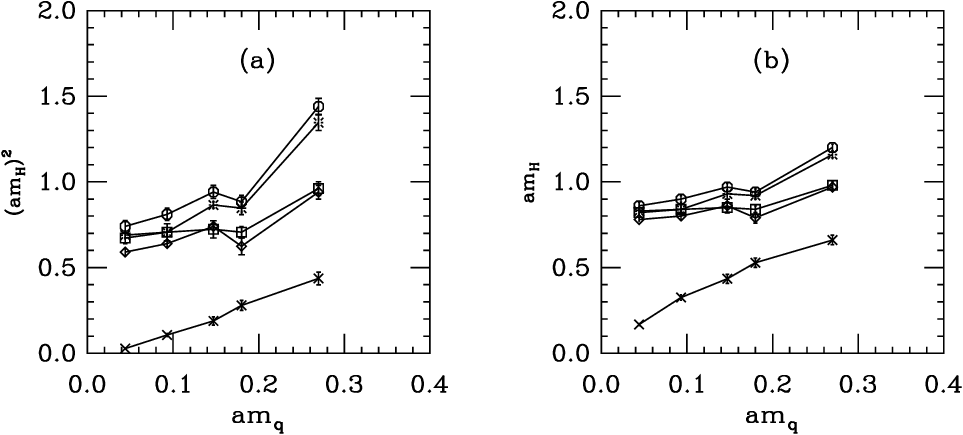}
\end{center}
\caption{Screening masses and $f_P$ for $\beta=6.0$ on volume $(12\times 8^2)\times 8$.
Labels as in Fig.~{\protect{\ref{fig:mpi5.51288}}}.
\label{fig:mpi6.01288}}
\end{figure*}

\begin{figure*}
\begin{center}
\includegraphics[width=.85\textwidth,clip]{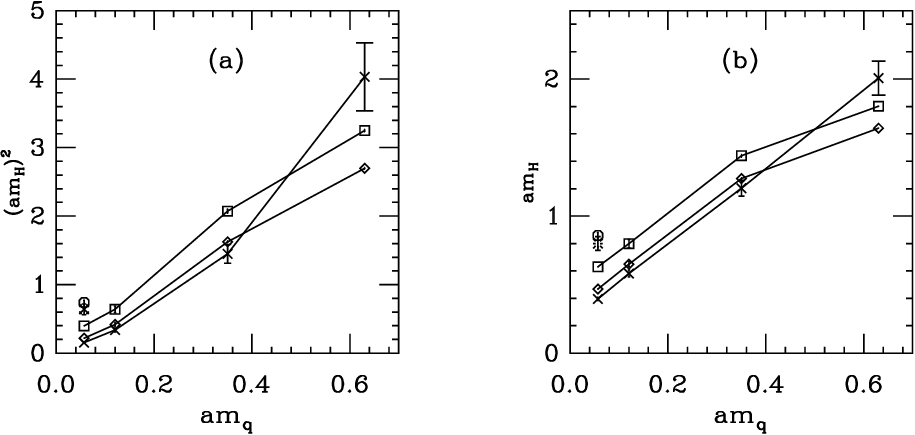}
\end{center}
\caption{Mass spectrum and $f_P$ for $\beta=5.3$ on volume $12^4$.
Labels as in Fig.~{\protect{\ref{fig:mpi5.51288}}}.
\label{fig:mpi5.31212}}
\end{figure*}

\begin{figure*}
\begin{center}
\includegraphics[width=.9\textwidth,clip]{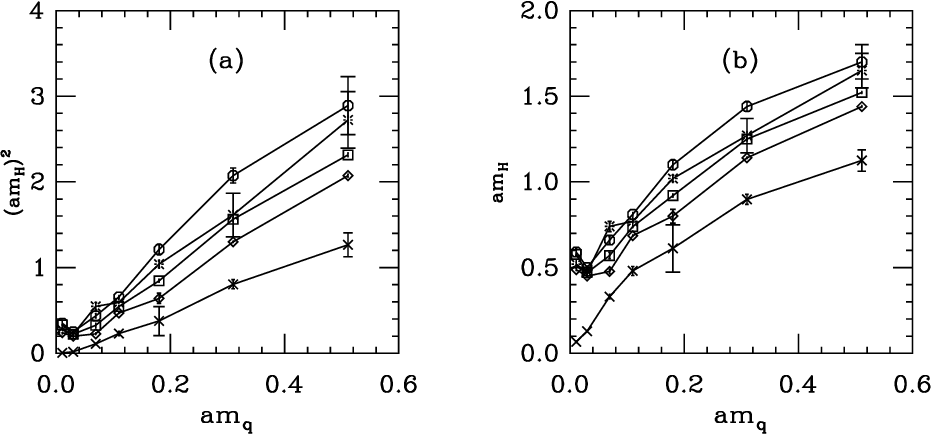}
\end{center}
\caption{Mass spectrum and $f_P$ for $\beta=5.7$ on volume $12^4$.
Labels as in Fig.~{\protect{\ref{fig:mpi5.51288}}}.
\label{fig:mpi5.71212}}
\end{figure*}

We illustrate this result with several sets of spectra
(Figs.~\ref{fig:mpi5.51288} and \ref{fig:mpi5.11288}--\ref{fig:mpi5.71212}).
In all cases the right panel displays a massive quantity (in lattice units, $am_H$)
 versus the AWI quark mass $am_q$, while the
left panel displays the squared quantity  versus $am_q$. In all graphs,  crosses show the
decay constant (with tadpole factor), while the other symbols are particle masses.
We have already presented Fig. \ref{fig:mpi5.51288}, data at $\beta=5.5$ on volume $(12\times 8^2)\times 8$.
The other finite temperature data sets produce similar behavior. Fig.~\ref{fig:mpi5.11288}
shows spectra from $\beta=5.1$, Fig.~\ref{fig:mpi5.21288}
shows spectra from $\beta=5.2$, and  Fig.~\ref{fig:mpi6.01288}
shows spectra from $\beta=6.0$.
The $\beta=5.1$ data set  shows a gap in the range of $m_q$ values which
comes of the abrupt change in $am_q$ with $\kappa$
(see Fig.~\ref{fig:mqvskappa}). Here the points at large
quark mass are confined, while in Figs.~\ref{fig:mpi5.21288} and~\ref{fig:mpi6.01288} all points are in
the deconfined phase. In all these cases,
the lightest Matsubara frequency is $\pi/8$ so the minimum meson mass 
according to Eq.~(\ref{eq:minmat}) would be about 0.8.

Scans of spectroscopy on $12^4$ lattices show
similar behavior to what we saw on the smaller lattices. Two examples are shown in
Fig.~\ref{fig:mpi5.31212} and
Fig.~\ref{fig:mpi5.71212}, $\beta=5.3$ and~5.7. At $\beta=5.3$ the three heaviest quark mass points are confined.
At $\beta=5.7$ the two heaviest mass points are confined.
The next point is on the transition and the rest are deconfined.
These are not screening masses, but measurements performed in the temporal direction---ordinary
spectroscopy. We have checked screening correlators at several of these points and they
produce identical results.

\subsection{Searches for curves of constant physics \label{sec:CCP}}
Motivated by experience with using lattice calculations for ordinary QCD phenomenology,
we attempted to find lines of constant physics, along which we might attempt to make
continuum predictions. We were not successful in doing this. We describe briefly what we tried to do,
and why it failed.

A curve of constant physics is a manifold of points in the space of all bare couplings 
on which a dimensionless ratio of two dimensionful parameters (with the same mass dimension)
is a constant. For our theory, this manifold is a line in the $\kappa,\beta$ plane.
Along that line, a lattice spacing can be defined from any dimensionful observable.
(For example, we might find a line of constant $m_{P}/m_V$, and along that line, $am_V$ 
would give the lattice spacing, at the fixed ratio.)

Our problem was twofold. 
First, we were squeezed between short-distance artifacts at small
$\beta$ and finite-volume effects at large $\beta$.  The former
became apparent when the potential became essentially linear at
all distances, meaning that the Sommer scale had fallen to the
neighborhood of the
lattice spacing.  The finite-volume effects become apparent, of course, at the deconfinement
transition in finite temperature and at the point where the linear
potential disappears at finite spatial volume.
As mentioned above, this left us with only a narrow strip in ($\beta,\kappa$)
lying along and beneath the confinement transition curve.

The second problem was that we have two kinds of observables, ``gluonic'' or ``confinement'' observables
(such as $r_0$, $\sigma$, and $T_c$) and ``mesonic'' or ``chiral'' observables (such as $m_P$, $m_V$ and
$f_P$). Curves of constant physics from confinement observables, such as lines of constant 
$r_0\sqrt{\sigma}$, simply did not look anything like curves from chiral physics, such as
lines of constant $m_{P}/m_V$.
Evidently there is no consistent way to define the lattice spacing in the region in question.

\section{Summary and future prospects  \label{sec:last}}

The two main issues addressed in this work are the existence of 
separation between chiral and confinement scales and the manifestation of a supposed infrared fixed point in the massless theory.
We have presented evidence against the former while the latter is
supported by
the behavior of the deconfinement transition curves as $N_t$ is changed.

Rather than seeing scale separation, we found only a single phase
boundary in the $(\beta,\kappa)$ plane for any given $N_t$.
The weak-coupling side of this phase boundary shows parity doubling and
a smooth connection to the region near $\Kc$, indicating that there is
no obstacle on the way to a chiral limit (or the Wilson-fermion
equivalent) without spontaneous symmetry breaking.
The smoothness of $f_P$ and its rapid decrease as $\Kc$ is approached
is consistent with this picture.

We did not see a chirally broken but deconfined phase.
Our study does not, however, rule out such a
phase near the would-be meeting point
of the deconfinement and $\Kc$ curves.  Here we were limited 
by the notorious numerical difficulties associated with unimproved
Wilson-clover fermions.
Moreover,
the lack of chiral symmetry in the Wilson formulation
(as opposed to staggered fermions
\cite{Kogut:1984sb,Kogut:1985xa,Karsch:1998qj}) 
leads to
the absence of a true order parameter, such as 
the quark condensate, that could be extrapolated to the massless limit.
After all, all of the evidence we know for scale separation
 involves measuring the condensate and the Polyakov loop, and seeing them order at different bare couplings.

Now let us expand on the significance of the $\Kth$ curves for the question of the IRFP.
Like all lattice gauge theories, our model possesses a strong coupling phase with a mass scale dictated by confinement.
This region must be separated from the basin of attraction of an IRFP because theories in this basin are conformal in the infrared, and thus cannot support a dynamically generated scale.
This basin, in turn, has to be a subset of
the $am_q=0$ (i.e., $\kappa=\Kc$) line, because a quark mass is a relevant coupling at a conformal fixed point.
The basin includes $\beta=\infty$ but it might not extend all the way to $\beta=0$; it can end at a (UV-attractive) critical point at some $\beta^*$ or at a first order bulk transition.

At finite volume---say, infinite spatial volume but nonzero
$N_t$---the system undergoes a confinement phase
transition or crossover; the phase boundary starts at infinite quark mass ($\kappa=0$) and extends into the diagram.  Let us suppose that it extends to an intersection with the $\kappa=\Kc$ line, so that the massless theory exhibits the phase transition as well.
(The alternative is that the phase boundary curves off towards $\beta=0$ at $\kappa<\Kc$.)
At infinite mass our theory is  a pure gauge theory so, as $N_t$ increases,
the transition will move to ever larger $\beta$. At $m_q=0$, however, the transition has to
remain in the confinement region of the infinite volume theory, {\em outside\/} of the basin of the IRFP.  This means that the deconfinement
lines for different $N_t$'s have to have an accumulation point on the $\kappa_c$ line qualitatively resembling
the situation shown in Fig.~\ref{fig:phase1}.  This is the $\beta^*$
critical point or the first order transition that bounds the IRFP's
basin.

A set of deconfinement lines that march up the $\Kc$ line, crossing the location of the putative IRFP,
would have been inconsistent with the existence of an IRFP. Similarly, observing chiral
symmetry breaking or confinement at parameter values near the IRFP would have been another inconsistency.
Our data point the other way.

SU(2) gauge theory with $N_f=2$ flavors of adjoint fermions is also
a candidate for walking technicolor, and may instead possess an IRFP \cite{Catterall:2008qk}.
This theory shares the main features
of our model. The SU(2) pure gauge theory possesses a second order confinement transition at finite $N_t$
and adjoint fermions preserve the associated $Z_2$ center symmetry; hence the pattern of finite temperature
transitions sweeping across the phase diagram at finite simulation volume is expected there, too.
If the massless theory has an IRFP, then the zero-mass limit of the confinement transition
must again terminate outside of its basin of attraction, and the phase diagram will resemble 
our Fig.~\ref{fig:phase1}.

We will return to the sextet theory with a study at smaller quark masses, which will become possible through the use of an improved lattice action.

\begin{acknowledgments}
We thank A.~Hasenfratz, N.~Christ, T.~G.~Kovacs, J.~Myers, and M.~Ogilvie for discussions.
B.~S. thanks D.~Dietrich, F.~Sannino, and the University of Southern Denmark for their hospitality at the Workshop on Dynamical Electroweak Symmetry Breaking in September, 2008, where many fruitful discussions took place.
This work was supported in part by the US Department of Energy and by the Israel Science Foundation
 under grant
no.~173/05.  Our computer code is based on version 7 of the publicly available code of the MILC 
collaboration~\cite{MILC}. 
\end{acknowledgments}


\end{document}